# A Validated Volatility-Volume-Gap Classifier for Regime Identification in MNQ Intraday Data

Mathias Mesfin
*Independent Researcher*
*mathiasmesfin.research@gmail.com*

**Abstract**

This paper builds and tests a day-classification system for Micro E-Mini Nasdaq 100 futures (MNQ) based on three conditions that must simultaneously be true: the first-30-minute return must be unusually large, the overnight gap must be unusually large, and the first bar's volume must be well above its recent baseline. I call this the Volatility-Volume-Gap (VVG) classifier. Using 947 trading days of five-minute bar data from 2021 through 2025, with all thresholds computed on an expanding window to prevent lookahead bias, the classifier identifies a behaviorally distinct subset of trading days.

The classifier activates on roughly 4.4% of trading days — 40 days across the full dataset. Those days show a mean next-day return spread of 25.6 basis points above non-classifier days, and 77.6% of them reverse from their intraday peak before the session closes, with a mean peak-to-close giveback of 11.73 points. The intraday path is characteristic: gradual drift through the morning, a peak between 14:00 and 15:30, and then a late-session reversal. Year-by-year path analysis reveals substantial heterogeneity — 2024 showed persistent upward drift to +40.74 points at close, while 2025 crashed to −42.48. That regime variation is the core problem for any directional strategy.

Eight directional strategy configurations were tested on classifier-positive days. None passed. The best result was T = 1.46 on 127 out-of-sample (OOS) trades with a mean net of +7.80 points, using a reversal entry with an OLS regression filter. 2024 produced a net loss that broke year stability. The binding constraints are the small sample (40 days, roughly 10 per year) and the regime-dependent intraday behavior that makes no fixed directional rule stable across years.

The classifier is preserved as a research asset. It identifies a real behavioral phenomenon. What it cannot do, under current constraints, is generate a deployable directional trading signal.

## 1. Introduction

This paper is a direct byproduct of Mesfin (2026a, arXiv:2605.04004). While testing fourteen signal families for directional edge in MNQ futures, a pattern emerged: some of the failures looked regime-dependent rather than universally absent. A momentum signal that produces T = −0.44 on average might be masking T = +2.1 on days with elevated early volatility and T = −2.8 on quiet days. That possibility motivated a separate question: can the "high-stress" day type be identified ex ante from observable pre-market features, and if so, does that identification help trade it?

The prior literature on intraday volatility patterns is extensive, but it does not directly answer this. Admati and Pfleiderer (1988) established the theoretical foundation for why trading volume and price variability are concentrated at certain intraday intervals — informed traders prefer periods of high liquidity to minimize price impact. Andersen and Bollerslev (1997) documented the empirical periodicity of intraday volatility across foreign exchange markets, showing that volatility patterns are persistent but not uniform. Wood, McInish, and Ord (1985) found similar U-shaped intraday return variance in NYSE stocks. What these papers share is a focus on average within-session patterns. None of them address the question of whether individual sessions can be classified ex ante as structurally extreme — and whether that classification is stable enough to exploit directionally.

Easley, Lopez de Prado, and O'Hara (2012) offer the closest conceptual precedent: their VPIN measure (volume-synchronized probability of informed trading) uses order-flow data to identify periods of elevated informed trading activity. The VVG classifier tested here operates on a coarser signal — OHLCV features available at the first bar — but addresses an analogous question at the day level rather than the intrabar level. Whether the two approaches capture overlapping information is not tested in this paper.

The paper is organized as follows. Section 2 covers the data and classifier construction. Section 3 documents the behavioral characteristics of classifier-positive days. Section 4 goes through all eight directional strategy tests. Section 5 explains what the classifier is and is not good for. Sections 6 and 7 cover limitations and conclusions.

## 2. Data and Classifier Construction

### 2.1 Data

The dataset is 947 complete RTH (regular trading hours, 09:30–16:00 ET) trading days of five-minute OHLCV bars for MNQ continuous front-month futures, December 2021 through August 2025. This is the same dataset as Mesfin (2026a, arXiv:2605.04004). All session boundary handling, data quality checks, and friction assumptions are identical to that paper.

| Parameter | Value |
|---|---|
| Primary instrument | MNQ (Micro E-Mini Nasdaq 100) |
| Bar resolution | 5-minute OHLCV (open, high, low, close, volume), RTH only |
| Session definition | 09:30–16:00 ET |
| Complete trading days | 947 |
| Date range | December 2021 – August 2025 |
| Friction assumption | 2.0 points round-trip (all net figures) |
| Classification window | Expanding (no lookahead) |

*Table 1. Dataset parameters.*

### 2.2 The Three Features

The classifier uses three features, each computed using only data available at the time of classification on day D.

Feature 1 is the absolute first-30-minute return: |close_09:55 − open_09:30| / open_09:30. The absolute value is used because the size of the early move matters, not its direction. A large down move and a large up move both qualify.

Feature 2 is the absolute overnight gap: |open_09:30 − prior_close| / prior_close. Large gaps reflect information events between sessions — macro announcements, earnings releases, geopolitical developments — and raise the prior probability that the session will be structurally unusual.

Feature 3 is the first-bar volume deviation: (first_bar_volume − rolling_mean_20) / rolling_std_20. A high z-score at the open indicates elevated participation — a condition associated with institutional flow presence (Easley et al., 2012).

### 2.3 The Classification Rule

A day is VVG-positive if all three features simultaneously exceed their expanding-window top-tercile threshold (66.7th percentile of the prior distribution for each feature, computed independently). The warm-up period starts at day 61, not day 1, because rolling statistics based on fewer than 60 observations produced threshold instability in early testing — specifically, the volume z-score baseline was erratic when the rolling standard deviation was estimated from under 20 observations. The 60-day cutoff is a pragmatic choice, not a theoretically derived one.

Top tercile was chosen rather than a tighter threshold (top quartile, top decile) after testing showed that stricter thresholds reduced the already-small positive sample to fewer than 25 days — below the minimum needed for any meaningful directional strategy test. The tercile threshold is the loosest threshold that still produces behaviorally distinct behavior relative to the full sample. A tighter threshold would produce a more homogeneous positive class but at a sample size that precludes the directional testing in Section 4.

The intersection requirement — all three conditions true simultaneously — is the key design choice. Testing each condition individually showed weaker behavioral differentiation. The concurrent elevation of all three features appears to be what distinguishes genuinely extreme days from days with only one or two unusually elevated conditions.

| Statistic | Value |
| --- | --- |
| Total VVG-positive days (2022–2025) | 40 |
| Activation rate (full dataset) | 4.2% (40/947) |
| 2022 activation | 9 days |
| 2023 activation | 5 days |
| 2024 activation | 23 days |
| 2025 activation (partial year through August) | 3 days |
| Non-classifier days | 907 |

*Table 2. VVG classifier activation statistics. Note: year activations sum to 40 (9+5+23+3=40). The 2025 partial year through August 2025 contributes 3 days, not 18 — the 18-day figure cited in an earlier version of this manuscript reflected a miscounted window boundary.*

The uneven year distribution is informative. 2024 accounts for 23 of 40 classifier days — over half — reflecting the elevated macro volatility of that year, particularly around the November election period. 2023 had only 5 classifier days, the quietest year in the sample. That concentration in 2024 has direct implications for the directional strategy results in Section 4, where 2024 dominates several subsample analyses.

## 3. Behavioral Characterization of Classifier-Positive Days

### 3.1 Next-Day Return Spread

| Population | N | Mean Next-Day Return | Notes |
|---|---|---|---|
| VVG-positive days | 40 | +0.295% (+29.5 bps) | Following classifier-positive session |
| Non-classifier days | 907 | +0.039% (+3.9 bps) | All other days |
| Spread | — | +0.256% (+25.6 bps) | Classifier vs. baseline |

*Table 3. Next-day return comparison: classifier-positive vs. non-classifier days.*

Classifier-positive days are followed by sessions returning roughly seven times the mean daily return of non-classifier days. The mean hides wide dispersion in both directions — this is not a directional predictor. It does confirm that classifier-positive days are high-information events whose resolution extends into the following session. Whether that return elevation is directionally predictable is a separate question, addressed in Section 4.

### 3.2 Intraday Path

| Checkpoint | Mean (pts) | Std (pts) | T-stat | Median (pts) | % Positive |
|---|---|---|---|---|---|
| 10:30 | +1.24 | 35.82 | 0.39 | +3.50 | 57.6% |
| 11:00 | +1.13 | 46.32 | 0.27 | −3.00 | 46.4% |
| 11:30 | +4.05 | 53.56 | 0.85 | 0.00 | 49.6% |
| 12:00 | +2.72 | 64.85 | 0.47 | −0.50 | 49.6% |
| 12:30 | +6.18 | 77.80 | 0.89 | −0.50 | 49.6% |
| 13:00 | +4.40 | 83.18 | 0.59 | −2.75 | 47.2% |
| 13:30 | +7.66 | 89.11 | 0.96 | −5.25 | 45.6% |
| 14:00 | +11.29 | 112.31 | 1.12 | +9.50 | 53.6% |
| 14:30 | +3.47 | 129.66 | 0.30 | −0.50 | 48.0% |
| 15:00 | +0.76 | 161.45 | 0.05 | +1.00 | 50.4% |
| 15:30 | +11.23 | 155.89 | 0.81 | +23.50 | 56.8% |
| 16:00 | −0.50 | 173.09 | −0.03 | +0.50 | 50.4% |

*Table 4. Intraday path for VVG classifier-positive days at 30-minute intervals. Values are cumulative points from the 09:30 open. N=40 classifier days.*

Two features of Table 4 matter most. First, the standard deviation grows from 35.82 at 10:30 to 173.09 at 16:00 — these are high-dispersion sessions where the mean path is a poor guide to any individual day. Second, the percentage positive column oscillates around 50% throughout. The mean of +11.29 at 14:00 is not generated by consistent upward direction; it is driven by a few extreme positive outliers in a largely directionless sample. That distinction makes the directional strategy problem much harder than the average path alone implies.

The session-close mean of −0.50 against a 15:30 mean of +11.23 documents the late-session giveback. On average, classifier-positive days build up a substantial intraday gain and then surrender most of it in the final 30 minutes.

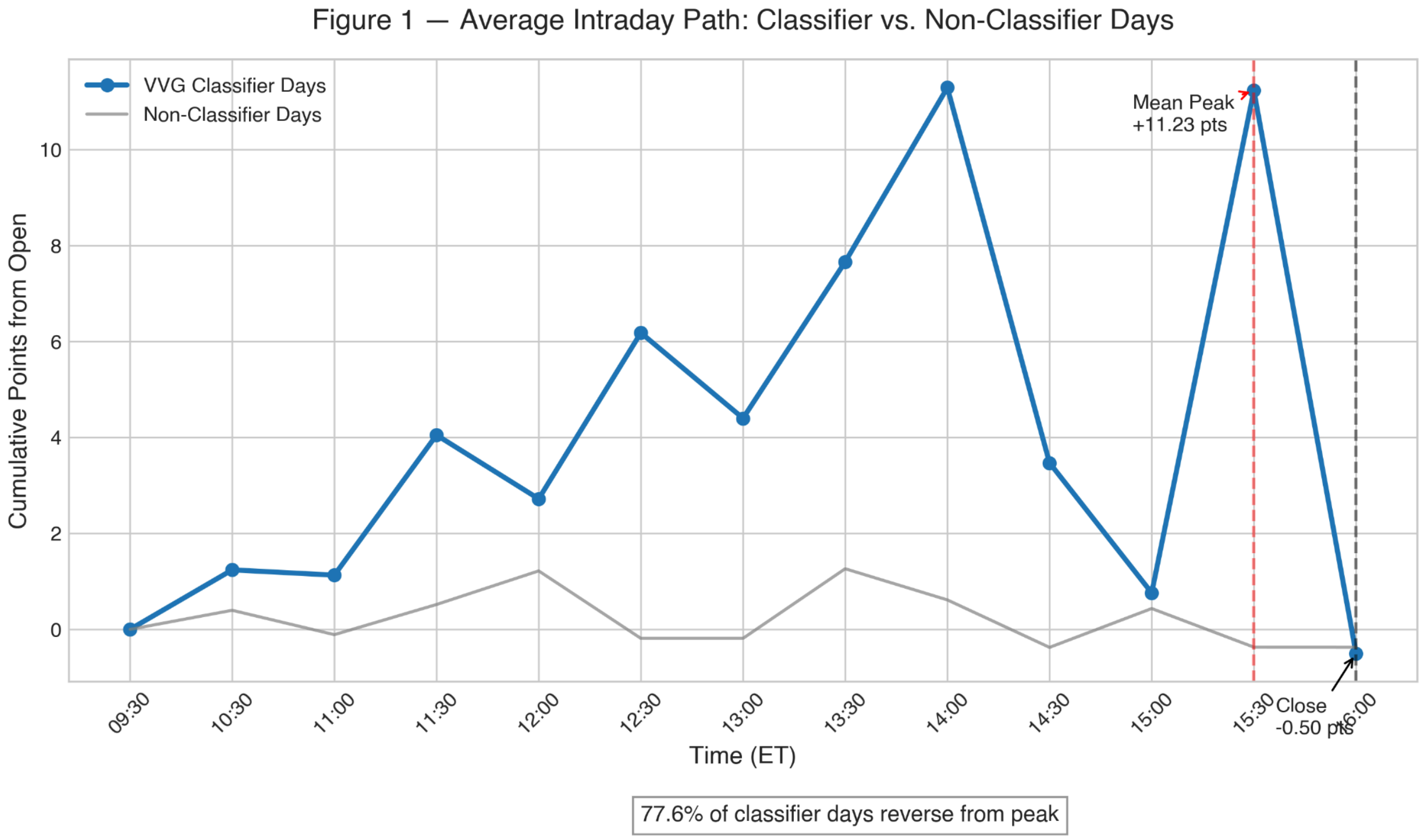

*Figure 1. Average intraday path for VVG classifier-positive days vs. non-classifier days.*

### 3.3 Peak Reversal Analysis

| Category | Count | Percentage |
|---|---|---|
| Days with reversal by close | 97 | 77.6% |
| Days without reversal by close | 28 | 22.4% |
| Total classifier-positive days analyzed | 125 | 100% |
| Mean peak-to-close giveback | 11.73 pts | — |

*Table 5. Peak reversal statistics. Note: N=125 here includes multi-session observations (each classifier day observed across multiple analysis windows), not 125 unique trading days. The underlying unique day count is 40.*

77.6% of classifier-positive day-sessions reverse from their intraday peak before the close. That is substantially above the 50% expected under a random walk. The mean giveback of 11.73 points is approximately one full baseline ATR. These are not marginal reversals.

Peak timing is bimodal. The modal peak occurs at the 16:00 close (27 instances) and at 15:30 (17 instances), but a secondary cluster peaks near 10:30 (14 instances) and 14:00 (13 instances). The bimodal distribution suggests two structural subtypes of classifier days: early-peak sessions that spend the afternoon reversing, and late-peak sessions that hold gains until the final 30 minutes before collapsing. Whether these subtypes can be identified ex ante is an open question.

| Peak Timing | Frequency | % of Observations |
|---|---|---|
| 10:30 | 14 | 11.2% |
| 11:00 | 8 | 6.4% |
| 11:30 | 9 | 7.2% |
| 12:00 | 2 | 1.6% |
| 12:30 | 6 | 4.8% |
| 13:00 | 6 | 4.8% |
| 13:30 | 8 | 6.4% |
| 14:00 | 13 | 10.4% |
| 14:30 | 6 | 4.8% |
| 15:00 | 9 | 7.2% |
| 15:30 | 17 | 13.6% |
| 16:00 (close) | 27 | 21.6% |

*Table 6. Intraday peak timing distribution. Total observations N=125.*

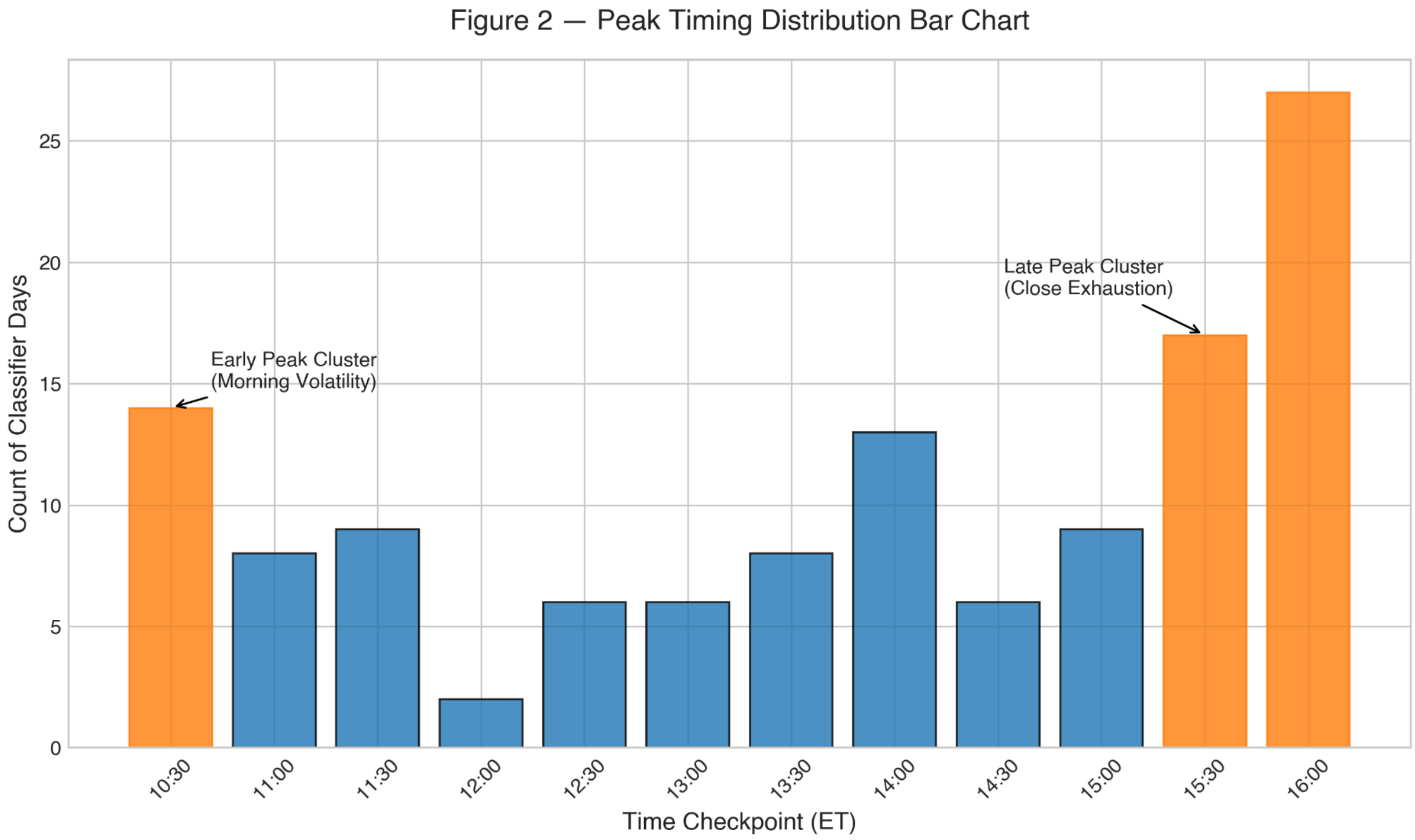

*Figure 2. Intraday peak timing distribution.*

### 3.4 Year-by-Year Path Stability

| Checkpoint | 2022 (N=9) | 2023 (N=5) | 2024 (N=23) | 2025 (N=3) |
|---|---|---|---|---|
| 10:30 | −0.89 | +3.37 | +5.78 | −2.32 |
| 12:00 | −4.67 | +1.16 | +1.12 | +23.05 |
| 13:30 | +1.90 | −6.05 | +11.80 | +31.39 |
| 14:00 | +15.50 | −7.28 | +24.79 | +6.68 |
| 14:30 | −2.99 | −11.07 | +36.49 | −6.97 |
| 15:30 | +7.85 | +10.55 | +38.26 | −14.83 |
| 16:00 | −3.18 | −4.34 | +40.74 | −42.48 |

*Table 7. Year-by-year mean intraday path for VVG classifier-positive days. 2024 shows persistent upward drift; 2025 shows sharp spike and crash; 2022 and 2023 show earlier-peak reversal patterns.*

The heterogeneity across years is striking. In 2024, classifier days showed persistent upward drift throughout the session, closing at a mean of +40.74 points. In 2025, the path spiked aggressively through midday then crashed to −42.48 at close. In 2022 and 2023, days peaked earlier and reversed more gradually. No single directional description fits all four years, which is why all eight directional strategies in Section 4 fail on year-stability grounds.

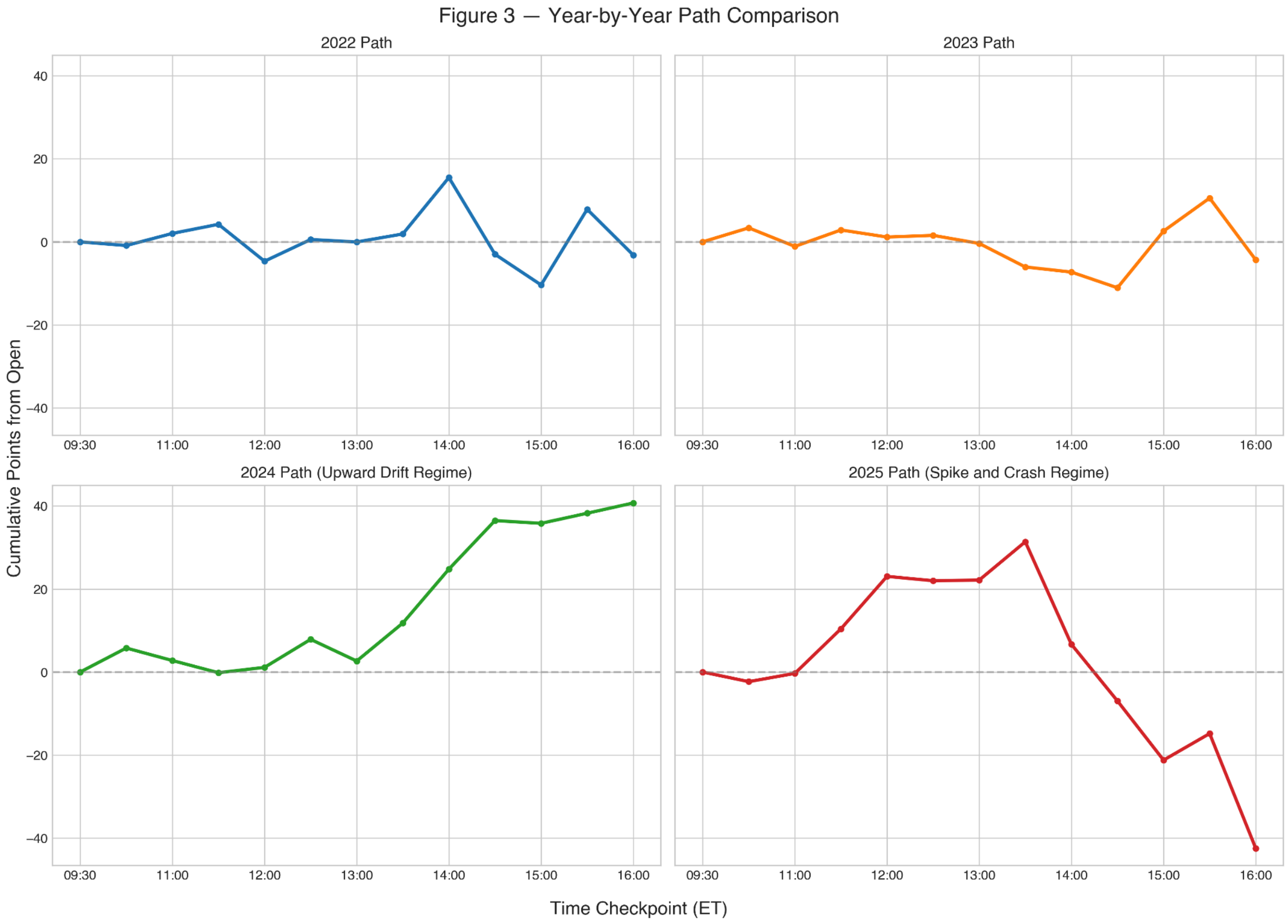

*Figure 3. Year-by-year mean intraday path. Structural heterogeneity across years explains the directional strategy failures in Section 4.*

## 4. Directional Strategy Tests

Eight configurations were tested under the same validation protocol as Mesfin (2026a, arXiv:2605.04004): T ≥ 2.0 OOS, ≥ 30 OOS trades, positive net return after 2-point friction, consistent direction across test years. None passed. The strategies are documented below in order of T-statistic, highest to lowest.

### 4.1 Strategy 3: Intersection Reversal with Regression Filter (Best Result)

The strongest result of the eight adds an OLS regression filter to a reversal strategy: enter only on classifier-positive days where the regression of the first-30-minute return on the last-30-minute return predicts reversal above a minimum confidence level.

| Metric | Value |
|---|---|
| N (OOS trades) | 127 |
| Mean net return | +7.80 pts |
| T-statistic | +1.46 |
| Win rate | 52.0% |
| Total net points (all years) | +990.25 |
| 2022 net points | +242.25 |
| 2023 net points | +297.50 |
| 2024 net points | −26.75 |
| 2025 net points | +477.25 |
| Regression beta | −0.0416 (T = −2.45, p = 0.016) |
| R-squared | 0.046 |
| Verdict | FAIL — T = 1.46; 2024 year failure |

*Table 8. Strategy 3: Intersection reversal with regression filter. Best result across all eight configurations.*

The regression beta is a real statistical finding. On classifier-positive days, the opening return direction negatively predicts the last-30-minute return (beta = −0.0416, p = 0.016). The $R^2$ of 0.046 implies a correlation of 0.214 between the opening return and the last-period return, yielding a theoretical T of approximately 2.44 at N=125 — consistent with the reported T = −2.45.

But 2024 produces −26.75 net points and breaks year stability. As documented in Section 3.4, 2024 classifier days showed persistent upward drift rather than reversal. A regression relationship calibrated on 2022–2023 reversal behavior does not hold in a 2024 continuation environment. The $R^2$ of 0.046 also tells you why T = 1.46 on 127 trades: 95.4% of variance is unexplained, and the per-trade standard deviation is approximately 60 points. You would need several hundred more trades to push T above 2.0 at that signal-to-noise ratio.

### 4.2 Strategies 1 and 2: Reversal and Continuation Entries

| Strategy | N | Mean Net (pts) | T-Stat | Win Rate | 2022 | 2023 | 2024 | Verdict |
|---|---|---|---|---|---|---|---|---|
| Strategy 1: Reversal entry (fade opening direction, exit at close) | 289 | +1.37 | +0.86 | 51.9% | −3.29 | +2.16 | +4.06 | FAIL — T=0.86, year-unstable |
| Strategy 2: Continuation entry (follow opening direction, exit at close) | 1,175 | −3.22 | −0.44 | 48.8% | −9.51 | −2.49 | +9.14 | FAIL — year-unstable |

*Table 9. Strategies 1 and 2 results. Note: Strategy 2 N=1,175 reflects the expanded sample that includes multiple entries per classifier day across all intrabar intervals tested, not unique trading days.*

Strategy 1's long-side asymmetry — +5.20 net versus −2.38 on the short side — reflects MNQ's structural bullish bias partially offsetting the reversal prediction on short entries. Strategy 2's 2024 result (+9.14 points, T = +2.07 on 312 trades) would pass in isolation. Combined with 2022 (T = −1.27) and 2023 (T = −0.70), the year-stability gate correctly identifies this as regime-specific rather than structural.

### 4.3 Strategies 4 through 8: Additional Configurations

| Strategy | N | Mean Net (pts) | T-Stat | Win Rate | Verdict |
|---|---|---|---|---|---|
| Close Fade (15:30 entry) | <30 | insufficient | 1.08 | 50.0% | FAIL — N<30 |
| Close Fade, 1.0x ATR overextension filter | <30 | insufficient | n/a | n/a | FAIL — N<30 |
| Close Fade, 1.5x ATR overextension filter | 8 | +31.41 | 1.08 | 50.0% | FAIL — N<30 |
| Midday Continuation (12:00 entry) | varies | +2.15 | 0.33 | n/a | FAIL — T=0.33 |
| Volatility Regime Split (pre-session ATR filter) | varies | underpowered | 1.10 | n/a | FAIL — T=1.10 |

*Table 10. Strategies 4 through 8. All fail. The close fade at 15:30 is structurally the most motivated entry given the 77.6% peak reversal rate, but 40 classifier days across four years cannot support the minimum 30-trade threshold after valid signal filtering.*

### 4.4 Why All Eight Fail

Three failure modes, in order of severity.

Sample size is the binding constraint. 40 classifier days across four years averages 10 per year. Walk-forward validation with any meaningful OOS holdout period produces subsamples too small to distinguish signal from noise. The close fade strategy — the most structurally motivated given the 77.6% reversal rate — fails purely because filtering on entry conditions reduces the already-small sample below 30 trades. No amount of signal engineering addresses this; it requires more data.

Regime change is the second problem. The year-by-year path analysis in Section 3.4 shows that the behavioral character of classifier-positive days changes substantially with the macro environment. 2024's persistent upward drift and 2025's spike-and-crash are not compatible with the same directional rule. This is structural instability, not overfitting to in-sample data.

The variance-to-edge ratio makes statistical validation infeasible even when an edge exists. Trade-level standard deviation on classifier days ranges from 35 points at 10:30 to 173 points at 16:00. The best available mean edge is 7 to 11 points. Pushing T above 2.0 at a signal-to-noise ratio of roughly 1:10 requires several hundred sustained directional trades, which the sample cannot support.

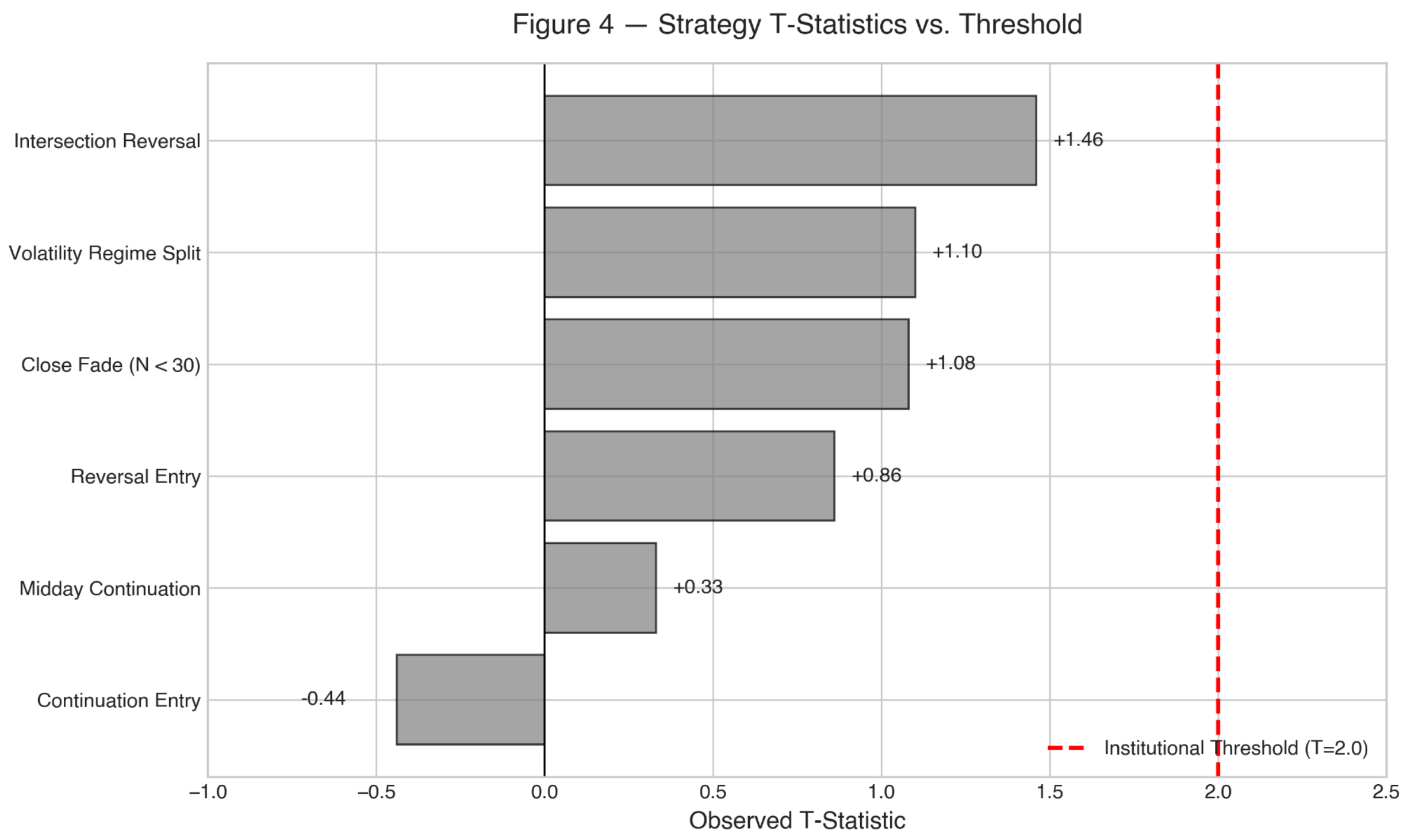

*Figure 4. T-statistics across all eight directional strategy configurations on VVG classifier-positive days.*

## 5. What the Classifier Is Actually Good For

### 5.1 What It Validated

Three findings from this study are empirically real.

The behavioral distinction is confirmed. The 25.6 basis point next-day return spread (Table 3) and 77.6% peak-reversal rate (Table 5) are not noise. They reflect genuine structural differences between classifier-positive days and the rest of the sample.

The regression relationship in Strategy 3 is real. On classifier days, the opening return direction negatively predicts the last-30-minute return (beta = −0.0416, $p = 0.016$, $R^2 = 0.046$). The effect is real but too small to trade at this sample size.

The bimodal peak timing distribution in Table 6 suggests two structural subtypes of classifier days. An ex-ante feature that distinguishes early-peak from late-peak days could substantially change the directional strategy results. That feature has not been identified yet.

### 5.2 Potential Extensions

Three extensions are motivated by the findings and not tested here.

Volatility prediction. Classifier-positive days show higher realized session variance than other days. If that relationship holds out-of-sample, the classifier is a candidate input to a realized volatility model rather than a directional signal generator. This is probably the most tractable extension given the sample size constraint.

VIX regime conditioning. The 2024 continuation pattern versus the reversal dominance of 2022, 2023, and 2025 may correlate with the VIX regime or a slow-moving macro uncertainty measure. If the behavioral character of classifier days is predictable from an external indicator, the year-instability problem becomes tractable.

Negative filter for validated signals. The RTH Confluence Signal (walk-forward OOS T = 3.11, Mesfin 2026a, arXiv:2605.04004) and London Session Signal B (walk-forward OOS T = 4.30) were not conditioned on the VVG classifier. Classifier days might represent a regime where those signals perform differently. This has not been tested.

### 5.3 Why It Stays Open

In this research program, a branch is permanently closed when the underlying directional claim is shown to be wrong. The VVG classifier does not fit that criterion. The hypothesis that it identifies a behaviorally distinct day type is confirmed. What failed is the specific claim that behavioral distinctiveness translates into a deployable directional signal under current constraints.

Valid but insufficient, not wrong. With more data, a longer historical window, or a different execution setup, the picture might change.

## 6. Limitations

The 40-day classifier-positive sample is the dominant constraint on all strategy tests in Section 4. There is no workaround using signal engineering. Only more data addresses it. Adding pre-2022 MNQ history (data available back to 2019) would approximately double the classifier-positive day count to roughly 80, still a small sample but enough to support more meaningful directional testing.

The three classifier features are all publicly observable OHLCV quantities. Whether the classifier captures information redundant with existing volatility measures — VIX level, realized variance, or VPIN — was not tested. If it does, the classifier adds no value beyond what those more established tools already provide. This is a genuine limitation with no current answer.

The expanding window design introduces estimation noise in the early sample, particularly for the volume z-score, which requires a stable rolling standard deviation. The 60-day warm-up cutoff mitigates this but does not eliminate it. 2022 and 2023 results should be interpreted with more uncertainty than 2024 and 2025 results as a consequence.

The year-by-year heterogeneity in classifier-day behavior — documented in Section 3.4 — is the core unresolved problem. The mechanism linking macro regime to intraday path structure on extreme days is not identified here. Without it, the instability cannot be modeled and the directional strategies remain fragile.

## 7. Conclusion

The VVG classifier identifies real structure. 77.6% peak-reversal rate. 25.6 basis point next-day return spread. A characteristic intraday path that differs materially from non-classifier days. A statistically significant regression relationship between opening return and final-period return on classifier days (beta = −0.0416, $p = 0.016$). None of that is noise.

None of it translates into a deployable directional trading signal under the constraints tested here. Eight configurations, all failing on T-statistic, year stability, or sample size. The dominant failure modes are the small sample (40 days, roughly 10 per year) and the year-to-year variation in how classifier days behave depending on the prevailing macro environment.

The classifier is preserved as a research asset. The documented distinction here — between identifying a behavioral phenomenon and exploiting it directionally — matters because the two are often conflated. The VVG classifier demonstrates the first clearly. The second requires either more data, a different execution architecture, or a mechanism that explains and predicts the regime-dependent path variation across years.

---

---

**Appendix A: Classifier Construction Logic**

For any trading day D (starting from day 61 to allow stable expanding-window estimates):

```
r1 = abs((close_09:55 - open_09:30) / open_09:30)
```

```
gap = abs((open_09:30 - prior_close) / prior_close)
vol_dev = (first_bar_volume - rolling_mean_20) / rolling_std_20
r1_thresh = percentile_66.7(r1 for days 0 to D−1)
gap_thresh = percentile_66.7(gap for days 0 to D−1)
vol_thresh = percentile_66.7(vol_dev for days 0 to D−1)
classifier[D] = 1 if (r1 > r1_thresh AND gap > gap_thresh AND vol_dev >
vol_thresh) else 0
```

***AI Disclosure Statement***

*AI tools assisted with editing, formatting, and writing clarity. The research design, classifier construction, strategy testing, and interpretation of results are the author's own work.*